\input stromlo

% redefine section to get rid of the break condition -- redefine
% to original at end of paper -- failure to do so leads to very
% very, very bad typsetting!!!!!!
\def\section2#1\par{\ifnum\secnum=0\medskip\maintextmode\fi
    \advance\secnum by 1 \bigskip
    \subsecnum=0
    \hang\noindent\hbox to \parindent{\bf\the\secnum.\hfil}{\bf#1}
    \smallskip\noindent}

\title Gravitational Lensing Limits On Early-Type Galaxies

\shorttitle Gravitational Lens Limits

\author C.S. Kochanek \& C.R. Keeton

\shortauthor C.S. Kochanek \& C.R. Keeton

\affil Harvard-Smithsonian Center for Astrophysics

\abstract Gravitational lenses are a unique new constraint on the structure
of galaxies.  We review the evidence that most lenses are early-type
galaxies, the optical properties of the lens galaxies, the evidence
against constant $M/L$ models, recent work on the axis ratios of
the mass distribution, and the role stellar dynamics plays in gravitational
lensing.  

\section2 Introduction

Gravitational lenses directly determine the transverse gravitational 
accelerations in a distant
galaxy or cluster.  Thus, they share the advantages of X-ray emission 
over stellar dynamics by avoiding all the additional assumptions needed to 
interpret line-of-sight velocities, and they 
can measure some properties of the lens galaxy with unprecedented accuracy. 
Nothing, however, is a panacea.
In particular, lenses are sensitive to gravitational perturbations along the 
line-of-sight between the observer and the source.  Rarity cannot be
considered a major weakness of gravitational lenses any longer, since the 
number of galaxy-dominated lenses is $\sim 25$ and 
continuing to increase.  In our brief review we will emphasize physical 
results relating to the structure of galaxies, and we will review neither
the physics of gravitational lensing (see Narayan \& Bartelmann (1996) 
or Schneider, Ehlers \& Falco (1992)) nor the
details of the observational data (see Keeton \& Kochanek 1996a).

Gravitational lenses constrain the properties of galaxies through their
statistical properties (numbers, image separations, morphologies etc.),
detailed models of individual lenses, and comparisons of the lens models 
to the optical properties of the lens galaxies and stellar dynamical models.
The issues we will discuss are (1) why we believe most of the lenses are
early-type galaxies; (2) the mass scale (or mass-to-light ratio) and the 
radial mass distribution of the lens galaxies; (3) the 
axis ratios of early-type galaxies; and (4) the interplay between
gravitational lensing and stellar dynamics.  

\section2 Lens Galaxy Colors and Luminosities

The mean image separation in a gravitational lens sample depends largely
on the typical velocity dispersion, 
$\Delta \theta \propto (\sigma/c)^2$, unless the numbers or properties of the 
lenses evolve rapidly at redshifts less than unity.  The numbers of lenses 
found in surveys depends on the cosmological model through
the comoving volume to the source $D^3$, the comoving density of lenses $n_*$,
and the square of the mean separation (the cross section), 
$N \propto n_* \Delta \theta_*^2 D^3$ (e.g. Turner, Ostriker \& Gott 1984,
Maoz \& Rix 1993, Kochanek 1996).  The $*$ denotes the values 
for a typical $L_*$ galaxy. Hence we can estimate the
relative numbers of spiral and early-type lens galaxies by the  
ratios of the products $ n_* \sigma_*^4$.  While the spirals are
more numerous ($n_e/n_s \sim 0.75 \pm 0.36$, Marzke et al. 1994), 
the early-types are significantly
more massive ($\sigma_e/\sigma_s \sim 1.5\pm0.15$), so we
expect most lenses will be early-types in the ratio of
about $4\pm2$.  
Spiral lenses have larger amounts of extinction and smaller
average image separations than early-type lenses, reducing their
detectability in optical and low resolution surveys respectively.

The optical data on lens galaxies are poor and inhomogeneous. Ground
based and WF/PC 1 images are limited by problems with resolution
and contrast relative to the images. Moreover,
many published lens magnitudes do not include the definition of the 
aperture used to determine the magnitude, and a
remarkably heterogeneous set of filters was used for the observations.
We hope to obtain a complete set of V, I, and H images of all
known lens galaxies in Cycle 7 to put the optical data on
a firmer footing, but in Keeton et al. (1996) we
attempt to systematically review the existing optical data.

Figure 1 shows the colors of the lens galaxies compared to standard
Bruzual \& Charlot (1993) evolution models.  It is difficult to
condense the results into a simple figure because no filter pair 
was used for more than two lens galaxies.  The lens galaxies
are generally as red or somewhat redder than expected for 
early-type galaxies.  One lens, MG~0414+0534, is far redder than
the models, and one lens, B~0218+357, is bluer.
B~0218+357 also shows HI absorption (Carilli et al. 1993), making it the best
candidate for a spiral lens in Figure 1. 

\figureps[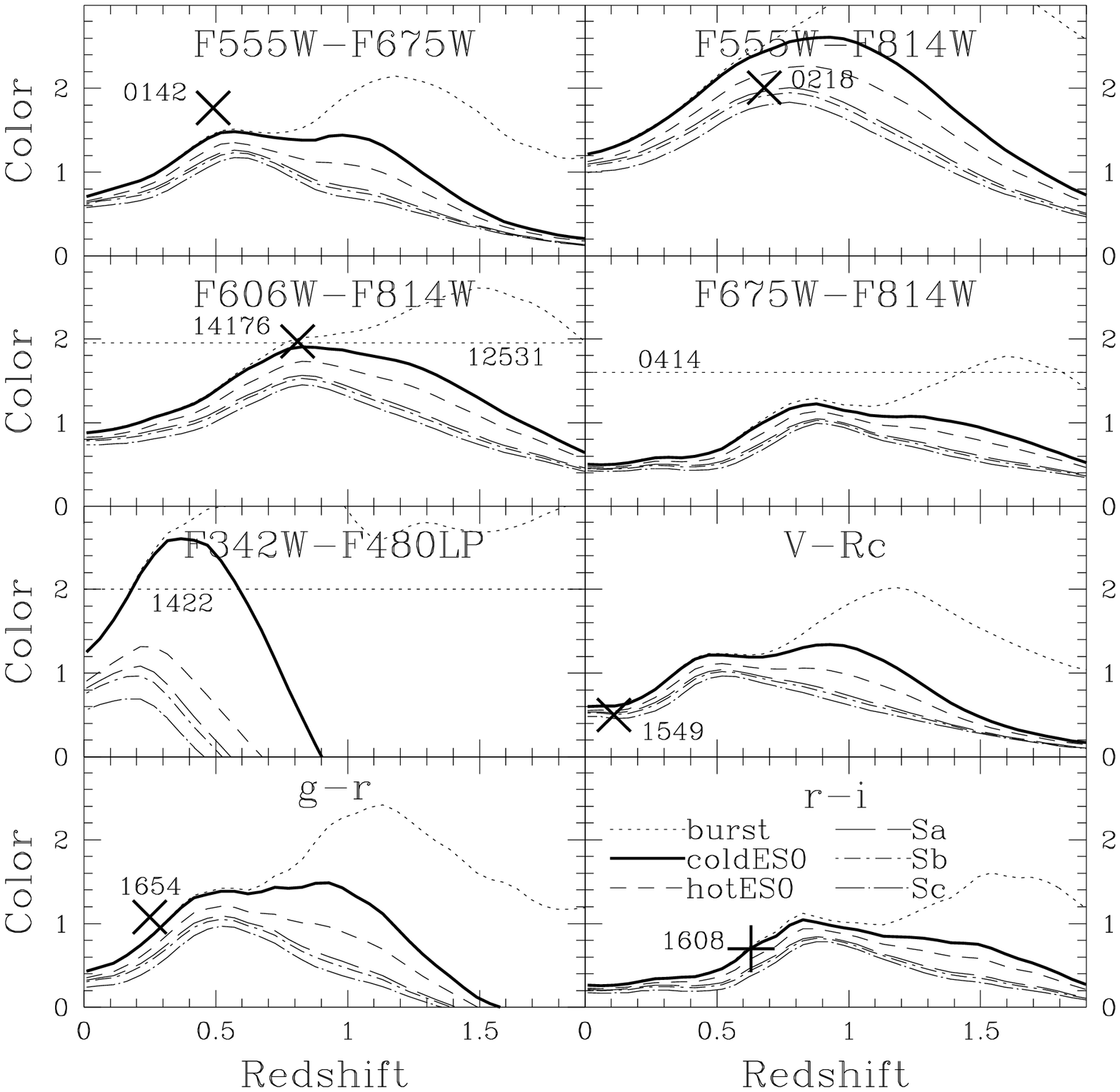,0.9\hsize] 1. Lens galaxy colors.  The crosses 
(known lens redshifts) and horizontal dashed lines (unknown lens redshifts) 
display the lens properties, and the curves are typical early-type to spiral 
color evolution models.  Typical uncertainties are 0.1 to 0.3 mag.

In Figure 2 we compare the ($z=0$) absolute B magnitude
of the lenses (using the evolution models to estimate the magnitude)
to the cosmology-corrected image separation (for $\Omega_0=1$),
to produce the lensing equivalent of the Tully-Fisher or
Faber-Jackson relation.  The problems with the lens magnitudes
are more severe here, and the typical uncertainties are probably
0.5 mag.  Nonetheless, there is broad agreement between the
observed and expected luminosities.  We treated B~0218+357 as
a spiral, and it would lie near MG~0751+2716 if
treated as an elliptical.  MG~0414+0534 again lies far off the 
locus of all other lens galaxies.  Its very
red spectrum is extremely peculiar, because most dusty galaxies
also have strong blue emission from ongoing star formation.

\section2 The Mass Scale and Radial Mass Density

Unlike the luminosity of the lens galaxy, the mass inside the
``ring'' defined by the lensed images is usually known to
$\sim$1\% for the four-image systems and radio rings and
to $\sim$10\% for the two-image systems.  The absolute scale depends
on the Hubble constant, weakly on the cosmological model
(usually 5--10\%), and weakly on potential perturbations along
the line of sight (also 5--10\% at worst).  Such accuracy for
a mass determination is unmatched by any other probe of galactic
structure except spiral galaxy rotation curves.

Maoz \& Rix (1993) established that de Vaucouleurs models 
with the van der Marel (1991) mass normalization of 
$(M/L)_{B*} = (10\pm 2)h$ are unable to produce the observed image 
separations, and Kochanek (1996) demonstrated that 
$(M/L)_{B*} = (20\pm4)h$ was required.
Most models of individual lens galaxies also have
projected mass-to-light ratios inside the ``ring''
defined by the images of roughly $20 h$ (also see Burke et al. 
1992), although the scatter and uncertainties are large because of the
optical photometry.
Figure 3 shows rough estimates of $(M/L)$ as a function of lens
redshift.  {\it All the uncertainty in Figure 3 is due
to the light rather than the mass!}  For example the PG~1115+080
point is a central (undefined) aperture magnitude rather
than an integrated magnitude. Given
improvements in the optical photometry it is a simple
matter to explore the evolution of the mass-to-light
ratio with lens redshift.  For comparison, if we assume a singular 
isothermal sphere is the prototypical dark matter model, the dark 
matter velocity dispersions required to fit the observed lens separations 
and the stellar dynamics of nearby galaxies agree with 
$\sigma_* \simeq 225 \pm 20 $ km s$^{-1}$ (Kochanek 1994, 1996),
and there is no evidence for the $(3/2)^{1/2}$ correction factor
introduced by Turner, Ostriker \& Gott (1984) (see \S5).   

\figuretwops[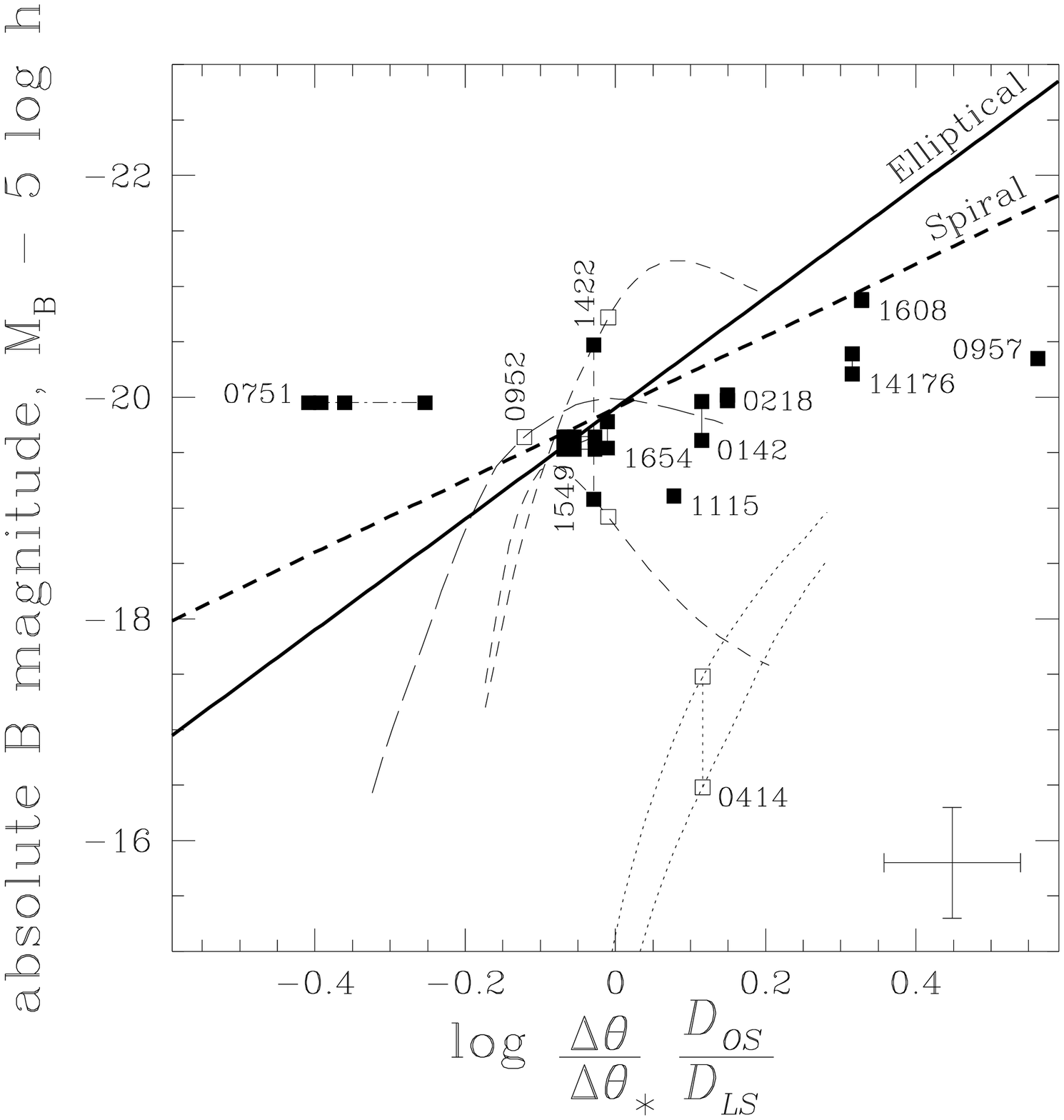,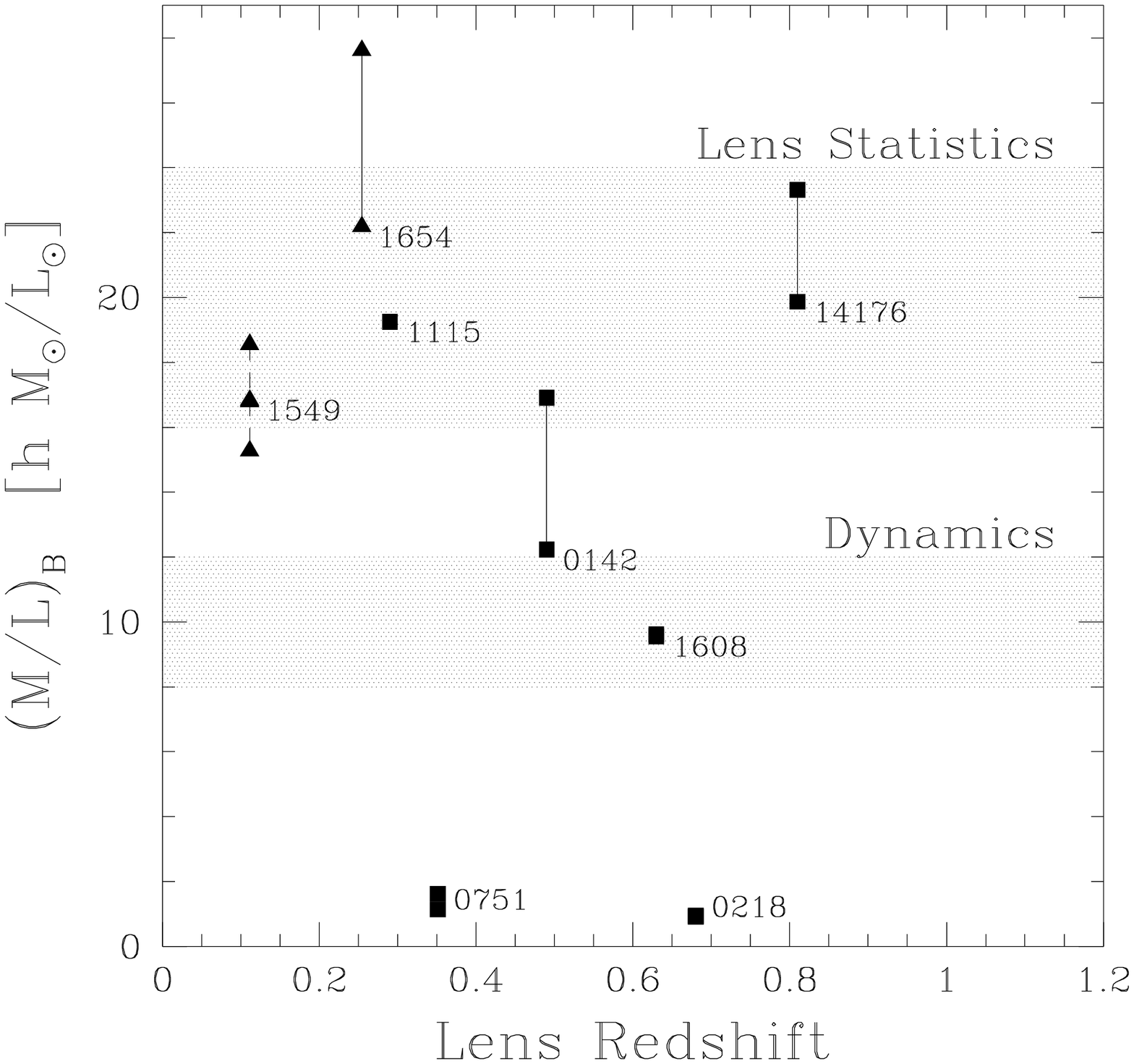,0.5\hsize] 2. Lens galaxy ``Faber-Jackson''
relation (left).  The estimated zero redshift absolute B magnitude of the lens
galaxies as a function of the cosmology-corrected image
separations.  The predicted correlations for standard models
of early-type (solid) and spiral (dashed) galaxies are shown by the
heavy lines.  The filled points mark lenses with known lens redshifts,
connected by a vertical line if we have estimates using more than one
color.  Filled points connected by a horizontal line (MG~0751+2716) show
the effects of not knowing the source redshift.  Empty points mark
the values for the most probable lens redshift where it is unknown,
and the curves cover the range of probable lens redshifts.  The
large error bar shows the level at which the curves and points can
be shifted relative to each other because of uncertainties in
$M_{B*}$ and $\Delta\theta_*$. \break
\noindent {\bf Figure 3}. Estimated B mass-to-light
ratios of lens galaxies corrected to $z=0$ (right).  
The scatter is the same as in Figure
2, but the switch to a linear scale highlights the problems
with the optical magnitudes.  For lenses marked by triangles
we have reasonable estimates of the aperture corrections,
while for the squares we do not.  Note that the probable
spiral lens B~0218+357 has a significantly lower $M/L$.

Directly determining the radial mass distribution is
significantly harder than determining the enclosed mass.
Kochanek (1991, also Wambsganss \& Paczy\'nski 1994) demonstrated 
that most lens geometries are insensitive to the radial 
mass distribution because all the images lie at a common 
radius from the center of the lens galaxy.
Only systems with images at significantly
different radii, or equivalently with extended radio
emission, can directly attack the issue of the radial
mass distribution.  The two published cases are the
MG~1654+134 radio ring (Kochanek 1995) and the 0957+561
VLBI double (Grogin \& Narayan 1996), both of which are
most consistent with singular isothermal
mass distributions.  We generally find that
many lenses weakly favor flat rotation curves and slowly
declining surface densities over more centrally concentrated
models.  

\figureps[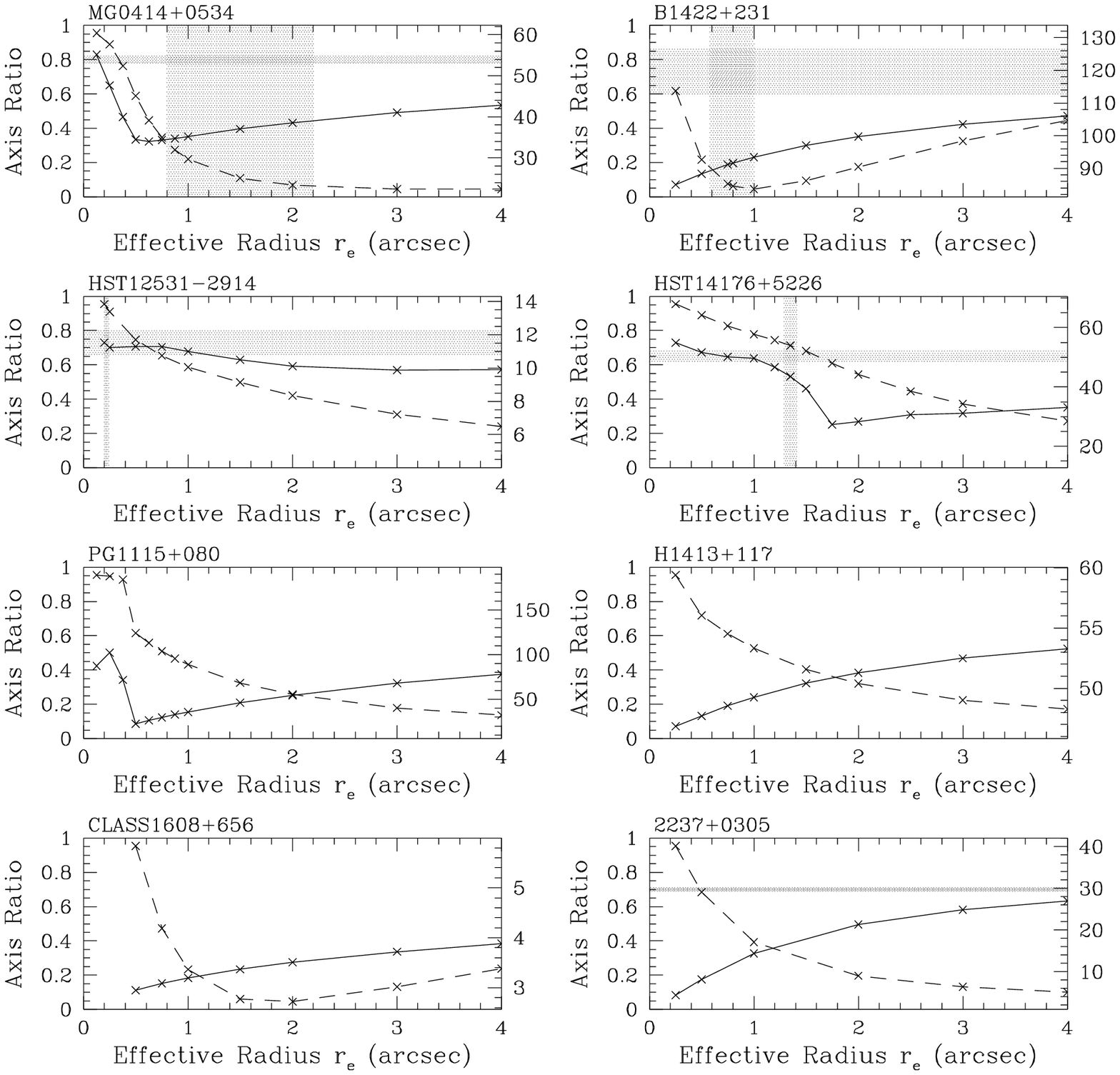,0.8\hsize] 4. $\chi^2/N_{dof}$ goodness-of-fit (right
scale, dashed lines) and model axis ratios (left scale, solid lines) for fitting 
the observed lenses with an ellipsoidal de Vaucouleurs model.  The grey bands 
mark the observational limits on the effective radius and axis ratio when known.

\section2 Ellipticities and External Perturbations

Most lens geometries accurately measure an ellipticity
and an orientation for the lens galaxy.  The orientation is
strongly constrained (to $\sim1^\circ$ in a four-image lens)
and independent of the monopole structure
of the lens, while the required shear depends on the monopole
structure.  The required ellipticity or shear
scales as $(1-\kappa_r)$ where $\kappa_r$ is the surface mass
density of the monopole at the radius of the ring in units of
the critical surface mass density for lensing (Kochanek 1991).  
Centrally concentrated models ($\kappa_r \simeq 0$) require 
twice as much ellipticity as isothermal models ($\kappa_r \simeq 1/2$).
The axis ratio differences between the centrally concentrated,
modified-Hubble models used by Nair (1996) and isothermal models 
are due to this scaling.

We have undertaken a survey to fit all the available lenses using
ellipsoidal de Vaucouleurs and softened power-law density
profiles ranging from isothermal through modified-Hubble and
Plummer models (Keeton \& Kochanek 1996b).   For example, Figure 4 
shows the $\chi^2$ and axis ratios for ellipsoidal de Vaucouleurs 
models of eight lenses as a 
function of the effective radius.  The model both fails 
to fit the data and requires an unusually flattened galaxy for
most of the lenses.  
If we switch to a singular isothermal ellipsoid,
the required lens galaxy ellipticities decrease and the
$\chi^2$ of the fits improve. Figure 5 
shows the best-fit axis ratios for a sample of four-image,
radio ring, and two-image lenses.  As expected from
theory, the four-image lenses are flatter than the radio rings
which are flatter than the two image lenses.

\figureps[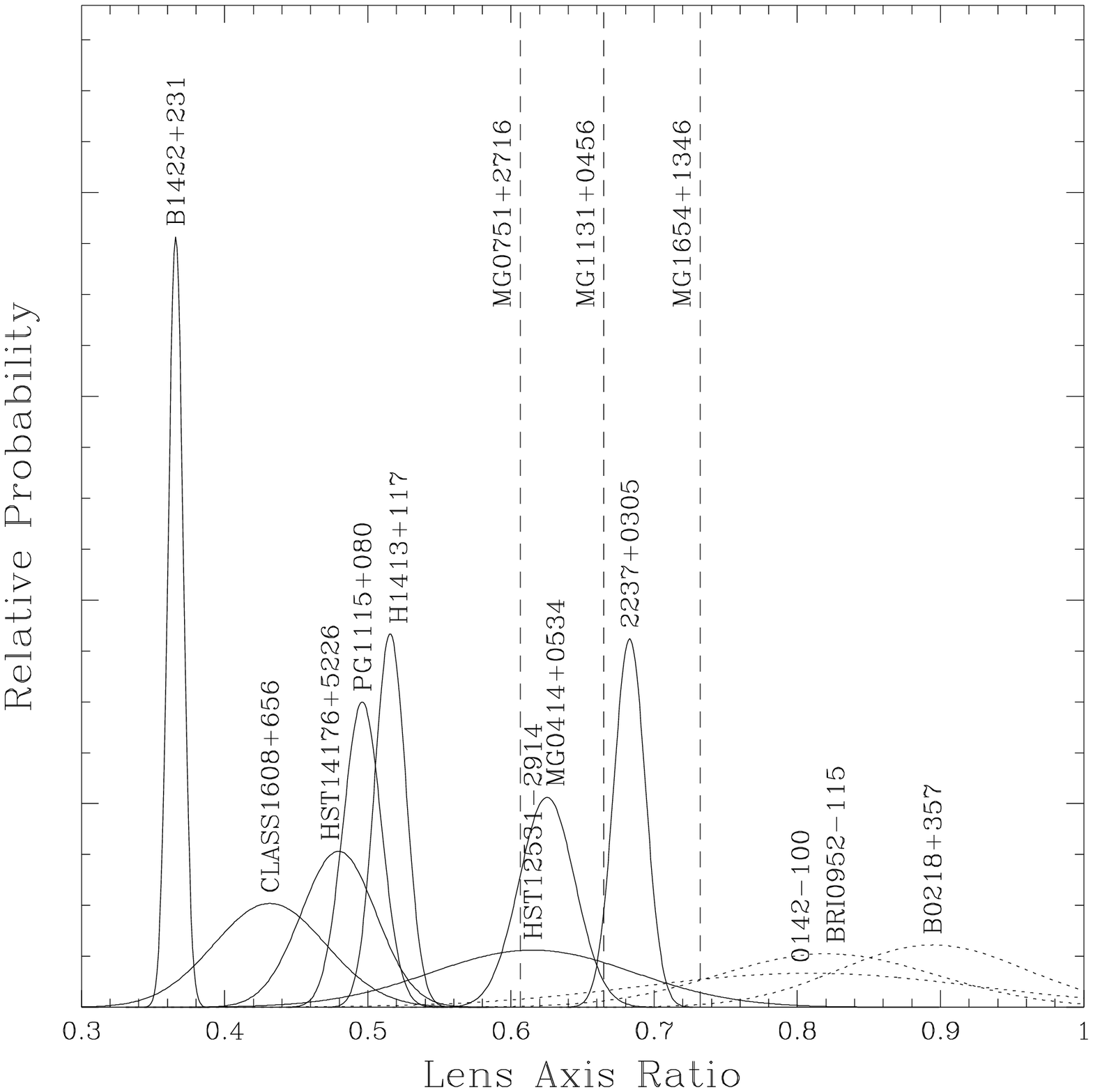,0.6\hsize] 5. Model axis ratios for singular
isothermal ellipsoids with four images (solid), radio rings
(vertical dashed lines), and two images (dotted).  We expect
the clear ordering of the axis ratios with morphology from
the cross sections.  These axis ratio estimates may be contaminated
by external tidal perturbations.  The axis ratios increase if we use
more centrally concentrated ellipsoids.

The relative numbers of different image morphologies also constrain
the axis ratio because the cross sections are functions of the 
ellipticity. For example, the four-image cross section is 
proportional to $\epsilon^2$.  Figure 6 shows the expected number 
of two-image, four-image, and three-image cusp lenses 
as a function of the axis ratio of a singular isothermal
ellipsoid.  The three-image cusp configuration has
three co-linear images offset to one side of the lens center
and has yet to be observed.  The observed sample has roughly
equal numbers of two-image and four-image systems, which requires
lenses with a mean axis ratio of nearly 2:1.  Ellipsoids are
much more efficient at producing four-image lenses than
external tidal perturbations, and as first noted by King \& Browne (1996)
the typical external shear perturbation $\gamma$ required to
fit the individual lenses woefully underpredicts the observed
number of four-image lenses.  Witt (1996) has also shown that some
lens galaxy positions are inconsistent with external shear as the
sole angular perturbation. The observed axis ratio distribution
for early-type galaxies, {\it including both true ellipticals and
the flatter S0s}, is statistically consistent with the distribution
needed to produce the observed numbers of four-image lenses and
to fit the individual lenses using singular isothermal ellipsoids.  
One problem we have yet to treat properly is the relative normalization of 
models with differing axis ratios (see \S5) --  
for a fixed amount of mass, flattened models
generally produce more lenses than spherical models.

\figureps[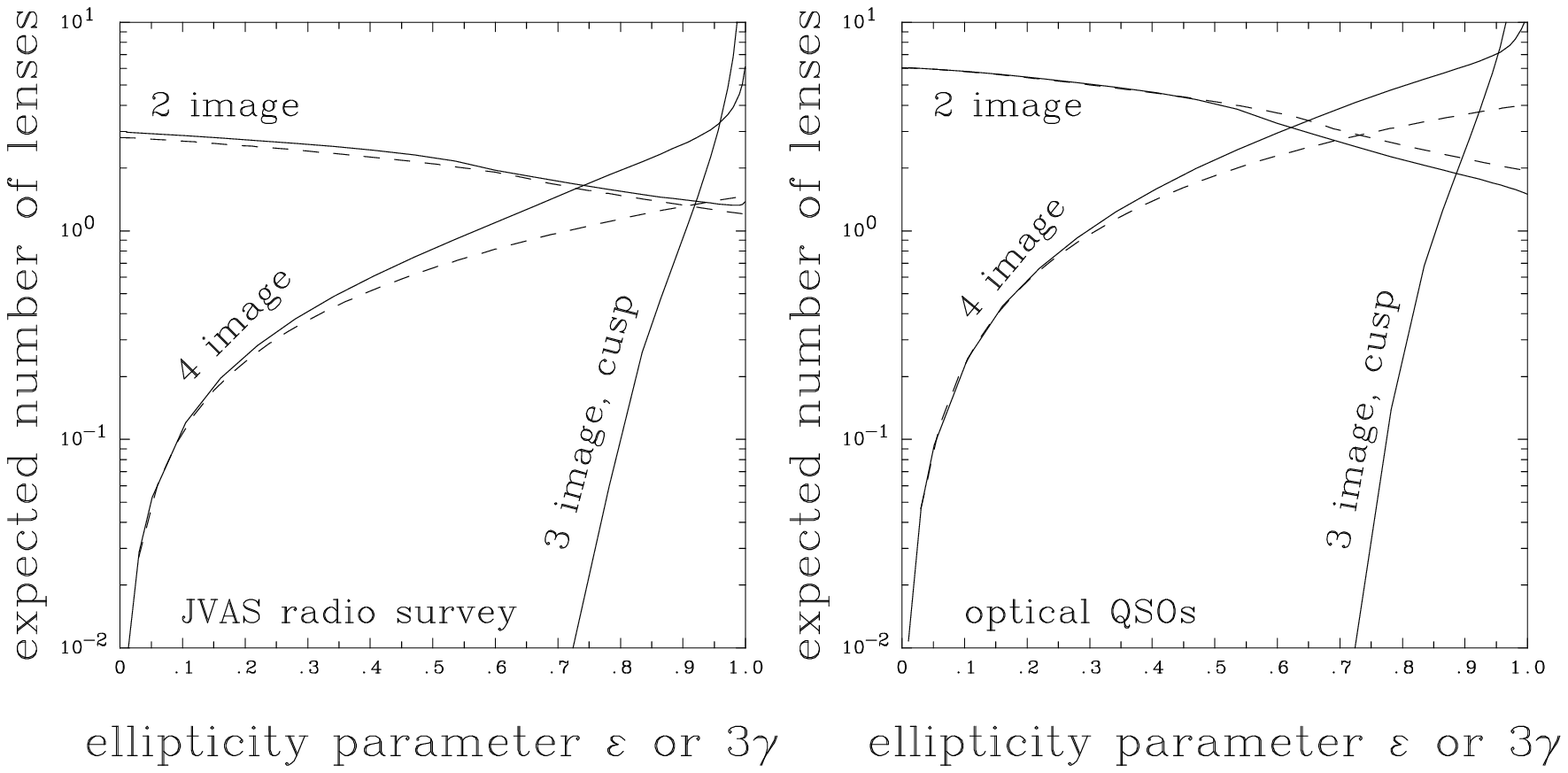,0.9\hsize] 6. Expected numbers of two-image, four-image
and three-image cusp lenses as a function of the external shear $\gamma$ (dashed)
or eccentricity $\epsilon$ (solid) for the JVAS radio lens survey (left, see
King \& Browne 1996)  
and optical QSO lens surveys (right).
The lens axis ratio is $r=(1-\epsilon)^{1/2}/(1+\epsilon)^{1/2}\simeq 1-\epsilon$
for singular isothermal ellipsoids.  

No lens we have examined, however, is well fit by a single ellipsoidal
density, but all lenses are well fit by the combination of an ellipsoid
and an independent external shear (Keeton, Kochanek \& Seljak 1996). The
dramatic improvement in the goodness of fit with the addition of a second
shear axis is a marked contrast to changes in the radial density distribution,
where  most lenses are equally well (or poorly) fit by radically differing
radial profiles (Kochanek 1991, Wambsganss \& Paczy\'nski 1994).
The origin of the second shear axis has yet to be fully understood, since
there are three possible sources.  First,
if the dark matter halo and the galaxy are aligned but have differing
triaxialities, then the projected dark matter can be misaligned from the
projected light (like the kinematic misalignment angle, e.g. Franx et al. 1991).
Second, the dark matter and the stars may not be in equilibrium, so that they
are intrinsically misaligned.  Third, there are external (tidal) perturbations
due to (in descending order of importance) galaxies, groups, and clusters 
correlated with the primary lens galaxy or near the line of sight (Kochanek \& 
Apostolakis 1988), and weaker perturbations due to large scale structure (Bar-Kana 1996). 
Several lenses (e.g. B~1422+231 (Hogg \& Blandford 1994) and PG~1115+080 (Schechter et al. 
1996)) appear to be part of a small
group of galaxies, where the other galaxies have the correct positions
to account for the secondary shear.
Surprisingly, we find that the external shear perturbations have
very little effect on the expected numbers of lenses even when they
are large enough to significantly alter the fits to the individual lenses.

The relative orientations of the major axes of the model and the stellar
distribution are a clean geometric test for the existence of both external
(tidal) perturbations and misalignments between the dark matter and 
the stars.  Unfortunately the available data are sparse and ambiguous.
In B~1422+231 the model and observed galaxy are aligned to $6^\circ \pm 15^\circ$
even though it is one of the strongest cases for the presence
of large external tidal perturbations (see Hogg \& Blandford 1994),
in MG~0414+0534 they are aligned to $8^\circ\pm 5^\circ$, in HST~12531--2914
they are aligned to $3^\circ\pm3^\circ$, and in HST~14176+5226 they
are aligned to $11^\circ\pm2^\circ$.  
Larger samples and reduced uncertainties in the galaxy orientations
are needed to determine the alignment distribution.  Whatever the
source of the second axis, the misalignments will be
larger for two-image lenses than four-image lenses.  Unfortunately, the
lens model alignments are less certain for the two-image systems unless
the lens galaxy position and image flux ratios are well constrained.
 
\section2 What Can Stellar Dynamics Do For Lensing?

Stellar dynamical models are important for normalization problems in
gravitational lensing.  For example, if we want to avoid using the observed
image separations to normalize the masses of galaxies, we must use existing
stellar dynamical estimates.  When doing so, it is important to use self-consistent
models and to avoid oversimplifications.  As we mentioned in \S3, an
oversimplified stellar dynamical model introduced by Turner, Ostriker \& Gott
(1984) implied that the dark matter dispersion $\sigma_{dm}$ for isothermal lens models
should be $\sigma_{dm} =(3/2)^{1/2}\sigma_c$ larger than the central velocity dispersion of the
stars $\sigma_c$.  Kochanek (1994) demonstrated that using the true luminosity 
profiles of galaxies (rather than an $r^{-3}$ power law) and the typical 
aperture sizes used in the dispersion measurements (rather than averaging over 
the entire galaxy) leads to $\sigma_{dm} \simeq \sigma_c$.  Since the
image separations scale as $\sigma_{dm}^2$ and the number of lenses
scale as $\sigma_{dm}^4$, the use of the poor dynamical model caused
an overestimate of the image separations by 50\%, and of the 
number of lenses by 125\%.  The resulting difference in the estimated 
cosmological matter density is $\Delta \Omega_0 \simeq 0.5$.

Oversimplifying the dynamical normalization can also bias inferences about other
parameters of the mass distribution.  For example, if we add a core radius $s$
to a dark matter distribution (e.g. $\rho \propto \sigma_{dm}^2/(r^2+s^2)$)  
then we must increase the velocity dispersion $\sigma_{dm}$ to compensate for
the reduced mass near the galactic center.  The singular model was normalized
to fit the observed central velocity distribution of the stars, and if we   
do not adjust $\sigma_{dm}$ upwards, the softened models will no longer fit 
the observations.  The correction to $\sigma_{dm}$ is small
(5--10\%), but it becomes a 10--20\% correction in the separations, and a
20-40\% correction in the expected number of lenses!  {\it These systematic
corrections are qualitatively and quantitatively important, and models that 
fail to include them will be wrong.}  With the addition of elliptical
structure, the normalization problem becomes considerably more difficult,
and we have yet to treat the problem in detail.  Finally, in 
0957+561 (e.g. Rhee 1991, Falco et al. 1996) there are measurements of 
the central velocity dispersion of the lens galaxy, which are used to
estimate the Hubble constant from the lensing time delay.  We must use 
dynamical models combining the mass profiles inferred from lensing with 
the luminosity profile of the galaxy to interpret the results 
(e.g. Grogin \& Narayan 1996).  Failure to include the  
full range of dynamical uncertainties (e.g. orbital isotropy)
in the models will lead to underestimates of the uncertainties in
the Hubble constant.

\section2 Summary

The optical properties of the lens galaxies are generally consistent
with most lenses being early-type galaxies with standard luminosities
and colors.  There are exceptions such as MG~0414+0534, which is
both anomalously red and faint, and B~0218+357, which is bluer, 
brighter and contains HI gas (i.e. a spiral).  
The heterogeneity of the data and imprecise aperture definitions 
make it difficult to compare lens galaxies to normal galaxies, but
the results semi-quantitatively agree with standard models. 
We desperately need to observe the lens galaxies using a uniform
set of filters and to carefully estimate their magnitudes, colors,
axis ratios, and orientations.  We could then, for example, track
the evolution of mass-to-light ratios with redshift.

Constant $M/L$ models appear to be are inconsistent with the 
lens data.  The value of $M/L$ needed to fit the image separations is 
double the stellar dynamical estimate, and ellipsoidal de Vaucouleurs models 
of individual lenses usually have large $\chi^2$ statistics
and large ellipticities.  Models with flat rotation curves normalized
to fit the image separations are consistent with stellar 
dynamics, are better lens models and have lower axis ratios.
Progress in determining the radial mass profile using lensing
depends on models of lenses with radial structure to the images.
Most of these lenses are radio rings, and we need deeper maps
with better visibility coverage to take full advantage of  
the rings as tools for probing the lens galaxies.

We have yet to understand the effects of external tidal
perturbations on the lens models.  External perturbations are
certainly present in several lenses (e.g. B~1422+231 and PG~1115+080),
but their amplitudes are hard to calibrate because many
effects of tidal perturbations can be mimicked by dark matter halos.
After all, if the radial mass distribution is dominated by dark matter
we should not be surprised if the angular mass distribution has a 
different axis ratio or triaxiality than the luminous matter.  
Galaxy formation simulations usually produce halos that are both
flatter and more triaxial than the visible stars (e.g. Warren et al.
1992, Dubinski 1992, 1994).
The distribution of misalignment angles between the major axes
of the lens models and the visible lens galaxies should be a good 
geometric probe of these problems.

The extreme sensitivity of lensing results to the mass scale of the
galaxies means that it is usually safer to determine the mass scale
using the observed image separations rather than stellar dynamical models.
If stellar dynamical models are used to set the masses, it is absolutely
critical to use self-consistent models.  Franx (see these
proceedings) has convincingly demonstrated that it is possible to 
measure velocity dispersions in galaxies 
comparable to the typical lens galaxy. We should try to determine
the central velocity dispersions of the brighter lens galaxies to
test and compare the lensing and stellar dynamical models.  

\acknowl  Acknowledgements:  We thank E. Falco, J. Hewitt,
  R. Narayan and P. Schechter for collaborations and discussions.
  CSK is supported by NSF grant AST-9401722, and CRK is supported
  by a National Defense, Science \& Engineering Fellowship. 

\references

Bar-Kana, R., 1996, ApJ, 468, 17

Bruzual, G.A., \& Charlot, S., 1993, ApJ, 405, 538

Burke, B.F., Leh\`ar, J., \& Conner, S.R., 1992, in Gravitational Lenses,
  eds., Kayser, R., Schramm, T., \& Nieser, L., (Springer: Berlin) 237

Carilli, C.L., Rupen, M.P., \& Yanny, B., 1993, ApJ, 412, 59

Dubinski, J., 1992, ApJ, 401, 441

Dubinski, J., 1994, ApJ, 431, 617

Falco, E.E., Shapiro, I.I., Moustakas, L.A., \& Davis, M., 1996, astro-ph

Franx, M., Illingworth, G., \& De Zeeuw, T., 1991, ApJ 383, 112

Grogin, N., \& Narayan, R., 1996, ApJ, 464, 92

Hogg, D.W., \& Blandford, R.D., 1994, MNRAS, 268, 889

King, L.J. \& Browne, I.W.A., 1996, MNRAS, 282, 67

Keeton, C.R., \& Kochanek, C.S., 1996a, in Astrophysical Applications of Gravitational Lensing, IAU 173,
  eds., C.S. Kochanek \& J.N. Hewitt (Dordrecht: Kluwer) 419

Keeton, C.R., Kochanek, C.S., \& Seljak, U., 1996, astro-ph/9610163

Keeton, C.R., Kochanek, C.S., \& Falco, E.E., 1996, in preparation

Keeton, C.R., \& Kochanek, C.S., 1996b, in preparation

Kochanek, C.S., \& Apostolakis, J., 1988, MNRAS, 235, 1073

Kochanek, C.S., 1991, ApJ, 373, 354

Kochanek, C.S., 1994, ApJ, 436, 56

Kochanek, C.S., 1995, ApJ, 445, 559

Kochanek, C.S., 1996, ApJ, 466, 638

Maoz, D., \& Rix, H.-W., 1993, ApJ, 416, 425

Marzke, R.O., Geller, M.J., Huchra, J.P., \& Corwin, H.G., 1994, AJ, 108, 437

Nair, S., 1996, in Astrophysical Applications of Gravitational Lensing, IAU 173,
  eds., C.S. Kochanek \& J.N. Hewitt (Dordrecht: Kluwer) 197

Narayan, R., \& Bartelmann, M., 1996, in the 1995 Jerusalem Winter School

Rhee, G., 1991, Nature, 350, 211

Schechter, P.L., Bailyn, C.D., Barr, R., et al., 1996, astro-ph/9611051

Schneider, P., Ehlers, J., \& Falco, E.E., 1992, Gravitational Lenses, (Berlin: Springer)

Turner, E., Ostriker, J.P, \& Gott, J.R., 1984, ApJ, 284, 1

van der Marel, R.P., 1991, MNRAS, 253, 710

Wambsganss, J., \& Paczy\'nski, B., 1994, AJ, 108, 1156

Warren, M.S., Quinn, P.J., Salmon, J.K., \& Zurek, W.H., 1992, ApJ, 399, 405

Witt, H.J., 1996, astro-ph/9608197

\bye